\documentclass{article}

   \usepackage[nonatbib, final]{neurips_2021}
\usepackage{amsmath}
\usepackage[utf8]{inputenc} 
\usepackage[T1]{fontenc}    
\usepackage{hyperref}       
\usepackage{url}            
\usepackage{booktabs}       
\usepackage{amsfonts}       
\usepackage{nicefrac}       
\usepackage{microtype}      
\usepackage{xcolor}         
\usepackage{graphicx}       
\usepackage[
  hyperref, backref, sortcites, backend=biber,
  date=year, sorting=nyt,
  isbn=false, doi=false, url=false,
  maxcitenames=2, mincitenames=1,
  maxbibnames=10, minbibnames=10,
  style=authoryear,
  citestyle=numeric-comp,
  dashed=true
]{biblatex}%
\makeatletter
\input{numeric.bbx}
\makeatother

\AtEveryBibitem{
    \clearfield{urlyear}
    \clearfield{urlmonth}
}

\title{Towards Lightweight Controllable Audio Synthesis with Conditional Implicit Neural Representations}

\author{%
  Jan Zuiderveld \\
  University of Amsterdam \\
  Royal Conservatoire The Hague  \\
  \texttt{janzuiderveld@gmail.com} \\
  \And
  Marco Federici \\
  AMLab\\
  University of Amsterdam \\
  \texttt{m.federici@uva.nl} \\
  \And
  Erik Bekkers \\
  AMLab\\
  University of Amsterdam \\
  \texttt{e.j.bekkers@uva.nl} \\
}

\begin{document}

\maketitle






%


\section{Introduction}


Controllable audio synthesis is a core element of creative sound design. Advancements in AI have made high-fidelity \textit{neural audio synthesis} achievable. However, the temporal resolution of audio and our perceptual sensitivity to small irregularities in waveforms make high quality audio synthesis a complex and computationally intensive task, prohibiting real-time, controllable synthesis within many approaches. In this work we aim to shed light on the potential of Conditional Implicit Neural Representations (CINRs) as lightweight backbones in generative frameworks for audio synthesis.

\textit{Implicit neural representations} (INRs) are neural networks used to approximate low-dimensional functions, trained to represent a single geometric object by mapping input coordinates to structural information at input locations. In contrast with other neural methods for representing geometric objects, the memory required to parameterize the object is independent of resolution, and only scales with its complexity. A corollary of this is that INRs have infinite resolution, as they can be sampled at arbitrary resolutions. Moreover, the fact that object samples are calculated independently allows sequential synthesis in memory- and/or computationally limited environments.

To apply the concept of INRs in the generative domain we frame generative modelling as learning a distribution of continuous functions \cite{dupontGenerativeModelsDistributions2021}. This can be achieved by introducing conditioning methods to INRs, e.g. by concatenating latent codes to input coordinates \cite{chenLearningImplicitFields2019} or modulating layer activations \cite{chanPiGANPeriodicImplicit2020}. We refer to this family of architectures as CINRs.

The dominant architectures in INR literature are multilayer perceptrons (MLPs) with traditional nonlinearities. These networks are biased towards low-frequency functions, known as \textit{spectral bias}\cite{rahamanSpectralBiasNeural}. Recently, several concurrent works have proposed to use periodic nonlinearities in INRs to facilitate learning finer, high-frequency details \cite{mildenhallNeRFRepresentingScenes2020, tancikFourierFeaturesLet2020, sitzmannImplicitNeuralRepresentations2020}. This increased expressiveness creates great potential for applying INRs in the domain of audio.

At the time of writing there is no published research applying CINRs to audio. To gauge the potential of CINRs for controllable audio synthesis we compare the ability of CINRs with periodic nonlinearities (PCINRs, based on $\pi$-GAN \cite{chanPiGANPeriodicImplicit2020}) and transposed convolution neural networks (TCNNs, based on WaveGAN \cite{donahueAdversarialAudioSynthesis2019}) to reconstruct different medium-scale sets of NSYNTH \cite{engelNeuralAudioSynthesis2017} waveform samples. 
 
Our experiments show that PCINRs learn faster and generally produce quantitatively better audio reconstructions than TCNNs with equal parameter counts. However, their performance is very sensitive to activation scaling hyperparameters. When learning to represent more uniform sets, PCINRs tend to introduce artificial high-frequency components in reconstructions. We validate this noise can be minimized by applying standard weight regularization during training or decreasing the compositional depth of PCINRs, and suggest directions for future research.\footnote{Sound examples and source code are publicly available at \href{https://janzuiderveld.github.io/audio-PCINRs/}{janzuiderveld.github.io/audio-PCINRs}.}

\begin{figure}[!htbp]
    \label{fig:PCINR_block}
  \includegraphics[width=1\linewidth]{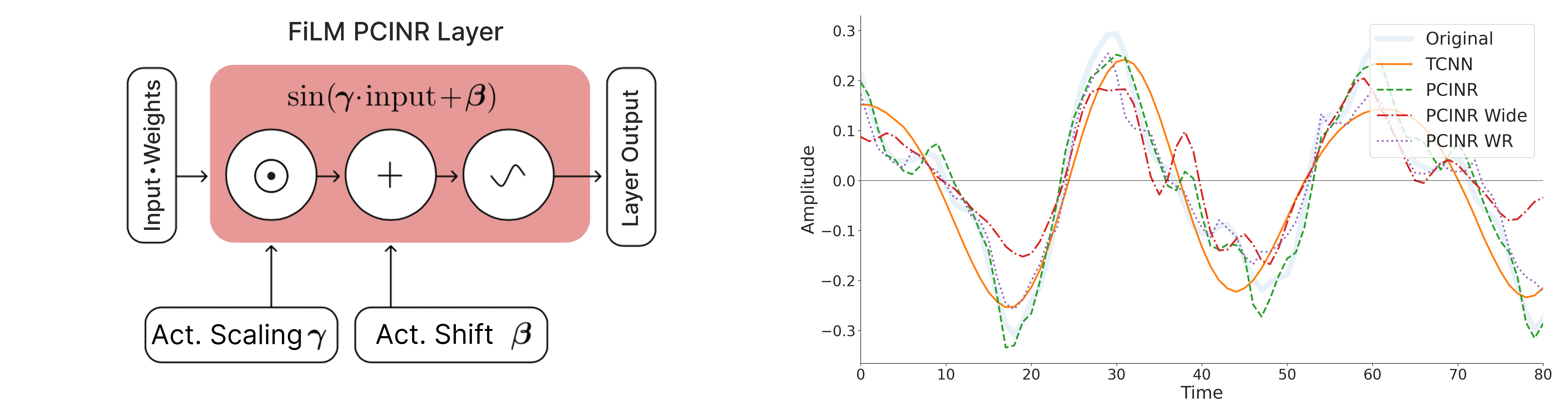}
  \caption{Left: Schematic overview of a FiLM PCINR Block. Adapted from Chan et al. \cite{chanPiGANPeriodicImplicit2020}. \\ Right: 80 samples (0.0005s) of a waveform of \textit{NSYNTH Diverse} and reconstructions of tested setups.}
  \label{fig:Waveforms_act_and_shape}
\end{figure}

\section{Overview of Methodology and Results} 
We investigate the ability of PCINRs to reconstruct two 1024 item sets of one second NSYNTH \cite{engelNeuralAudioSynthesis2017} waveform samples, \textit{NSYNTH Keyboard} and \textit{NSYNTH Diverse}. The performance of TCNNs with equal parameter counts is reported as a baseline. Tested PCINRs are conditioned using feature-wise linear modulation (FiLM \cite{perezFiLMVisualReasoning2017}). In our FiLM implementation, latent codes modulate the activation scaling and shift before periodic nonlinearities are applied. Preliminary experiments showed high-frequency noise in the output of PCINRs. We test two regularization methods to minize this. \textit{PCINR Wide}, an MLP with fewer layers and more hidden neurons. \textit{PCINR WR}, the same MLP with weight regularization (WR) applied during training. See Figure \ref{fig:Waveforms_act_and_shape}. Details in Section \ref{method} and \ref{exper}.   

Three quantitative evaluation metrics are reported: Contrastive learning-based multi-dimensional Deep Perceptual Audio similarity Metric (CDPAM \cite{manochaCDPAMContrastiveLearning2021}), Multi-resolution Short-Time Fourier Transform Mean Squared magnitude Error (Multi STFT MSE) and Mean Squared Error (MSE). Details in Section \ref{sec:evaluation}. Experimental results are reported in Table \ref{tab:results}.



\begin{table}[!htbp]
  \centering
  \caption{Means of CDPAM, Multi STFT MSE and MSE for NSYNTH Diverse and NSYNTH Keyboard over all set items. Standard deviations are calculated over 3 training runs. For all metrics a lower score is better. Silence and noise ($\mu=0$, $\sigma=1$) scores are reported for reference.} 
  \label{tab:results}
  \begin{tabular}{lllllll}
  \toprule
  {} & \multicolumn{2}{l}{CDPAM: Overall quality} &  \multicolumn{2}{l}{Multi STFT MSE: Noise} & \multicolumn{2}{l}{MSE: Absolute similarity} \\
  {Arch.} &                Diverse &     Keyboard &              Diverse &      Keyboard &               Diverse &     Keyboard \\
  \midrule
  TCNN  &            0.43 $\pm$ 0.21 &  0.43 $\pm$ 0.27 &                  \textbf{0.06} $\pm$ 0.05 &  0.06 $\pm$ 0.04    &  4.05        $\pm$    0.42 & 8.67 $\pm$  0.83      \\
  PCINR   &  \textbf{0.35} $\pm$ 0.24 &  0.48 $\pm$ 0.18 &          \textbf{0.06} $\pm$ 0.02 &  0.14 $\pm$ 0.04   &  \textbf{1.32}   $\pm$  0.3 & 4.48    $\pm$  1.02            \\ 
  - Wide      &            0.56 $\pm$ 0.28 &   \textbf{0.36} $\pm$ 0.2 &             0.11 $\pm$ 0.05 &  0.08 $\pm$ 0.02   &  6.55 $\pm$  0.16 &     1.64 $\pm$  0.03         \\
  - WR      &            1.11 $\pm$ 0.05 &  0.71 $\pm$ 0.05 &        0.07 $\pm$ 0.0 &   \textbf{0.04} $\pm$ 0.0      &   4.53  $\pm$  0.22 &     \textbf{0.85}  $\pm$  0.1          \\ 
  \midrule
  Silence             &            1.65 $\pm$ 0.0 &  1.77 $\pm$ 0.0 &         0.16 $\pm$ 0.00 &  0.16 $\pm$ 0.00 &     46.19 $\pm$  0.0 &     78.74 $\pm$  0.0  \\
  Noise         &            1.06 $\pm$ 0.27 &  1.06 $\pm$ 0.32 &     2.27 $\pm$ 0.06 &  2.32 $\pm$ 0.04   &   $\sim 1\times10^3$ &   $\sim 1\times10^3$   \\
  \bottomrule
  \end{tabular} 
\end{table}

\section{Overview of Conclusions and Future work}
PCINRs exhibit exceptional expressiveness making them well-suited for modelling sets of high-frequency one-dimensional continuous functions such as audio. They perform superior with TCNNs in quantitative metrics representing quality, noise and similarity. Qualitatively, PCINRs model more details, but induce a certain amount of high-frequency noise in the output signal. They offer direct control for balancing this trade-off between expressiveness and local smoothness. However, the optimal hyperparameters and performance of PCINRs are very sensitive to dataset characteristics such as note- and timbral diversity. It is unknown if these observations hold for large-scale PCINRs.




An idea which looks promising in light of our results, yet unexplored in INR literature, is splitting PCINRs fully-connected MLP in parallel subnetworks. In Section \ref{future} we argue this would reduce the propagation of noise induced at the recurrent stationary points in periodic nonlinearities, and by doing so minimize noise in the learning signal during training and in the output signal during synthesis. This argumentation is supported by multiple research directions \cite{liVisualizingLossLandscape2018, zhangDeepNeuralNetworks2018, haeffeleGlobalOptimalityNeural2017, fortDeepEnsemblesLoss2020}. 

\section{Methodology}
\label{method}
 
\subsection{Problem formulation}
\label{sec:fromal_problem}

We are interested in learning to represent a set of continuous audio waveforms covered in dataset $D$ consisting of $N$ discretely sampled waveforms. Waveforms in datasets $D$ are represented by point sets $X_{i}$, consisting of $M$ equally-spaced sampled amplitude values from the corresponding continuous amplitude functions $\mathit{y}_{i}: \mathbb{R}^{1} \rightarrow \mathbb{R}^{1}$:

$$
    \mathcal{D}=\left\{X_{i}\right\}_{i=1}^{N}, \quad X_{i}=\left\{\left(t_{j}, a_{j}\right): a_{j}=\mathit{y}_{i}\left(t_{j}\right)\right\}_{j=1}^{M}.
$$

\noindent Where $t_{j}$ are time coordinates, and $a_{j}$ are the corresponding amplitudes or air pressure measurements at these time coordinates.

We represent audio waveforms by directly approximating amplitude functions $\mathit{y}_{i}$ with continuous functions $\mathit{f}_{i}: \mathbb{R}^{1} \rightarrow \mathbb{R}^{1}$, parameterized by MLPs $\phi_{i}$ with sets of (learnable) parameters $\theta_{i} \in \Theta$.


We assume parameter sets $\theta_{i}$ live in a low-dimensional manifold. We define:
\begin{itemize}
    \item A function mapping latent representations $\mathbf{z}$ to intermediate latent representations $\mathbf{w} \in \mathcal{W}$,  $\mathbf{w} = g(\mathbf{z})$ with $g:\mathcal{Z} \rightarrow \mathcal{W}$, parameterized by a neural network $\psi$ with learnable parameters $\xi$, also known as the latent mapping network.
    \item A set of parameters $\alpha$ functioning as weights in $\phi_i$ shared between representations. 
    \item A function defining how intermediate latent representations $\mathbf{w}$ influence shared parameters $\alpha$ to obtain $\theta_{i}$, $\theta_{i} = c(\mathbf{w}_{i}, \alpha)$, also known as the conditioning method.
\end{itemize}

\noindent The shared parameters $\alpha$, latent mapping network $\psi$ and conditioning method $c$ together parameterize the conditional distribution $p(\theta|\mathbf{z})$, allowing us to map latent representations to the parameter space $p(\theta) = E_z[p(\theta|\mathcal{Z}=\mathbf{z})]$. By learning $\mathbf{z}_{i}$ during training using an autodecoder setup, obtaining $\mathbf{w}_{i}$ by mapping $\mathbf{z}_{i}$ through $g(\mathbf{z}_{i})$,  conditioning shared parameters $\alpha$ using $\mathbf{w}_{i}$ and conditioning method $c(\mathbf{w}_{i}, \alpha)$, we obtain parameter set $\theta_{i}$ for $\phi_{i}$. Then, we can evaluate $\phi_{i}|\theta_{i}$ at time coordinates $t_j$ to obtain corresponding amplitude approximations $\hat{a_j}$:
$$
\quad \theta_{i} = c(g(\mathbf{z}_i), \alpha), \quad \hat{a_j} = \phi_{i}(t_j|\theta_{i}).
$$
 
\noindent We optimize $\phi_{i}|\theta_{i}$ to represent functions $\mathit{y}_{i}$ with absolute fidelity, MSE.

\subsection{Architectural details}

\paragraph{PCINR, based on $\boldsymbol{\pi}$-GAN}
\label{subsec:Baseline Decoders pi-GAN}
PCINR is adapted from the $\pi$-GAN \cite{chanPiGANPeriodicImplicit2020} decoder, which is designed to produce implicit 3D radiance fields conditioned on $\mathbf{z}_{i}$. Ignoring the final parallel layer which integrates ray directions, $\pi$-GAN parameterizes $\phi_i$ as an 8 layer MLP with sine nonlinearities and a constant amount of hidden units (256). The conditional distribution $p(\theta|\mathbf{z})$ is parameterized as follows:

\begin{itemize}
    \item Latent mapping network $\psi$ consists of a 3 layer MLP with ReLU activations.
    \item The conditioning mechansism $c$ falls under the category of FiLM. $\mathbf{w}^\gamma$ and $\mathbf{w}^\beta$ vectors are shared between layers, drastically decreasing the total size of $\mathbf{w}$.
    \item $\mathbf{w}^\gamma$ does not scale layer activations directly, but is summed element-wise with the predefined activation scaling factor $\omega_0$ we know from SIRENs, resulting in the following conditioning method $c$ in layer $k$:\footnote{Technical descriptions in the paper of $\pi$-GAN seem to indicate that $\mathbf{w}_k^\gamma$ directly replaces $\omega_{0, k}$. This resulted in instable behaviour in preliminary experiments. E. Chan shared parts of their implementation showing the discussed conditioning method in which $\mathbf{w}_k^\gamma$ and $\omega_{0, k}$ are summed.} \\
    $$     \theta^{b}_k, \theta_{k}^{\omega_0} = c(\mathbf{w}, (\boldsymbol{\alpha}^{b}_k, \omega_{0, k})) = \mathbf{w}_k^\beta + \boldsymbol{\alpha}^{b}_k, \mathbf{w}_k^\gamma + \omega_{0, k}.
    $$
    \item Shared parameters $\boldsymbol{\alpha}$ populate all weights native to $\phi_i$.
\end{itemize}

\begin{figure}
    \centering
    \includegraphics[width=0.7\textwidth]{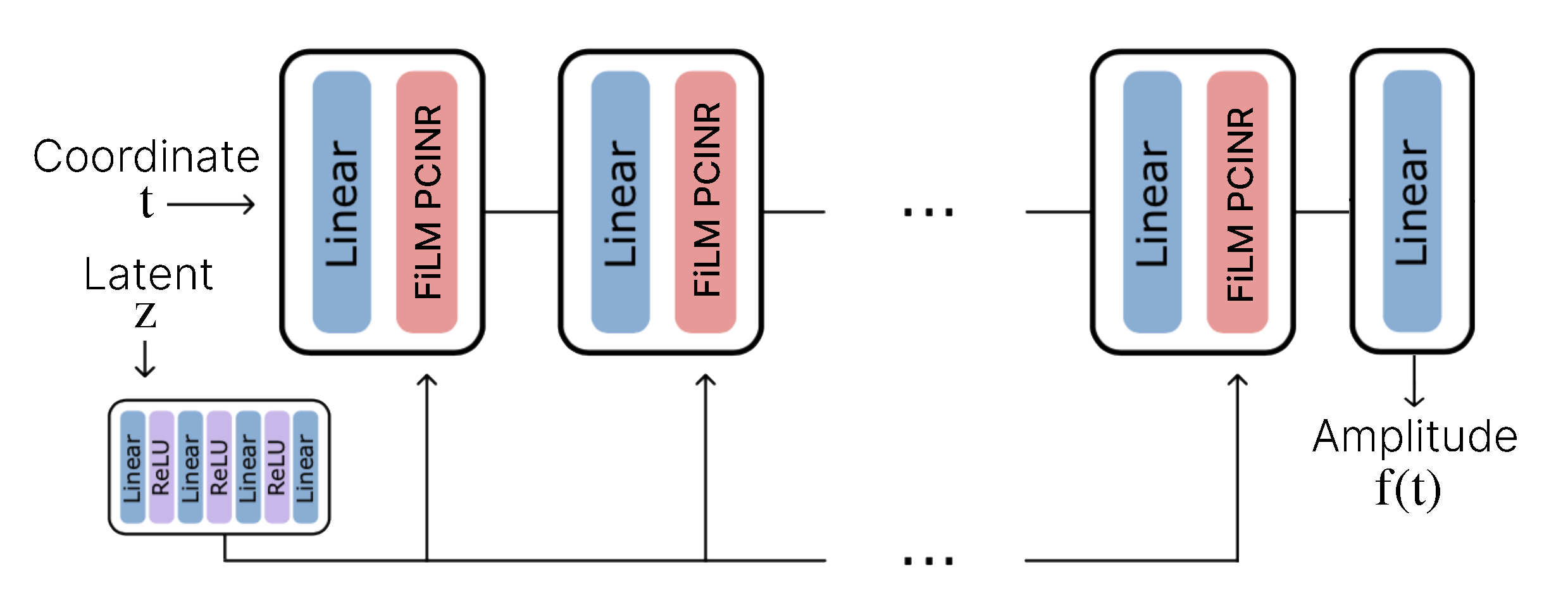}
    \caption{PCINR network structure overview. See figure \ref{fig:PCINR_block} for a schematic overview of the inner workings of a FiLM PCINR Block. Figure is adapted from Chan et al. \cite{chanPiGANPeriodicImplicit2020}.}
    \label{fig:pi-gan-arch}
\end{figure}

See figure \ref{fig:pi-gan-arch} for a schematic overview of the network structure.

\paragraph{TCNN, based on WaveGAN}

\label{subsec:Baseline Decoders WaveGAN}
TCNN is adapted from the WaveGAN decoder, which is based on of DCGAN \cite{radfordUnsupervisedRepresentationLearning2016}'s decoder, which popularized usage of GANs for image synthesis. This decoder uses transposed convolution to iteratively upsample low-resolution feature maps into a high-resolution image. In the WaveGAN decoder this is modified to work with audio by replacing its two-dimensional 5x5 filters with one-dimensional filters of length 25, and changing the stride factor for all convolutions from 2x2 to 4. These changes result in WaveGAN having the same number of parameters, numerical operations, and output dimensionality as DCGAN. Because DCGAN outputs 64x64 pixel images — equivalent to just 4096 audio samples — one additional layer is added to the model resulting in 16384 samples, slightly more than one second of audio at 16 kHz.

\begin{figure}[!htbp]
    \centering
    \includegraphics[width=0.7\textwidth]{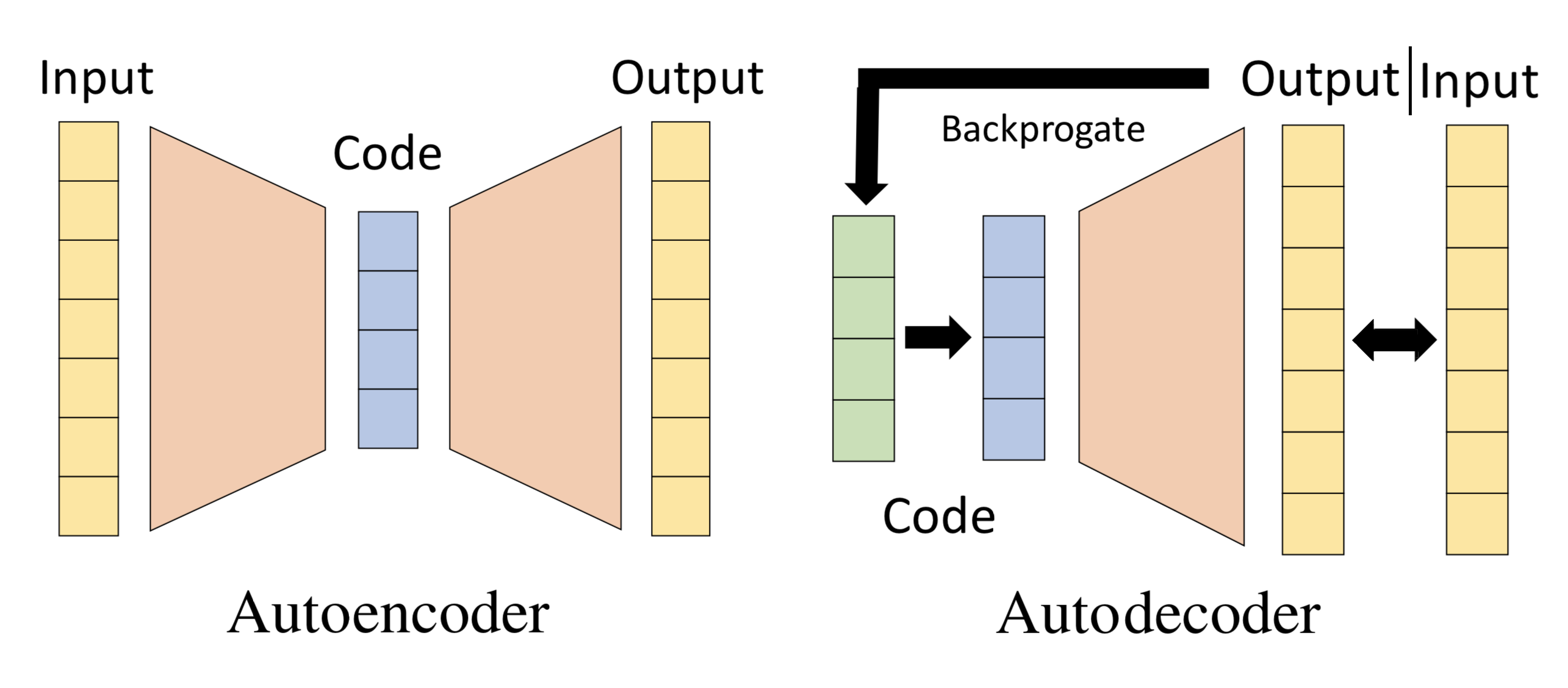}
    \caption{Schematic overview of autoencoder and autodecoder architectures. Figure adapted from Park et al. \cite{parkDeepSDFLearningContinuous2019}}
    \label{fig:autodecoder}
\end{figure}

\paragraph{Latent code inference} Autodecoders were introduced recently by Park et al. \cite{parkDeepSDFLearningContinuous2019}. They have no encoder network. Latent embeddings $\mathbf{z_{i}}$ are instead treated as learnable parameters rather than inferred from observations at training. By storing and updating intermediate latent embeddings $\mathbf{z_{i}}$ with backpropagated training errors during training the decoder can function as an encoder, see Figure \ref{fig:autodecoder} for a schematic overview. 

Due to the direct optimization of latent embeddings $\mathbf{z_{i}}$, autodecoders generally require less training iterations to reconstruct datasets faithfully. Autodecoders are more parameter efficient by not having an encoder, resulting in less operations per iteration and lower memory requirements during training. Finally, autodecoders can alleviate incompatibility issues between encoders and decoders. This is especially useful when dealing with implicit neural representations (INRs), as these are less trivial to be mirrored to use as an encoder.

\paragraph{Feature-wise linear modulation (FiLM)} FiLM proposes to apply element-wise\footnote{In the case of convolutional networks, FiLM is applied feature map wise.} scaling and shifting of intermediate layer activations based on latent representations $\mathbf{z}_i$. Framing the method in the conceptual framework described in section \ref{sec:fromal_problem} we note the following specifics:
 
\begin{enumerate}
    \item Intermediate latent representations $\mathbf{w}$ have the dimensiononality of twice the total amount of hidden units of conditioned layers in $\theta_i$, one half for scaling ($\mathbf{w}^\gamma$) and one half for shifting ($\mathbf{w}^\beta$) layer activations.
    \item The conditioning function $c$ is applied as shown in the equation below. Note that parameters controlling bias $\theta_{i}^{b}$ are replaced by $\mathbf{w}_{i}^\beta$.
    \item All other parameters $\theta$ in $\phi_i$ are parameterized by $\alpha$, shared between representations. 
\end{enumerate}

$$
y_{k}(\mathbf{x}_k)=act_k(\mathbf{w}^\gamma_k \cdot \mathbf{W_k}\mathbf{x}_k + \mathbf{w}^\beta_k).
$$

FiLM is a unification of methods that modulate intermediate layer activations, as in normalization layers introduced originally in batch norm \cite{ioffeBatchNormalizationAccelerating2015}. Other implementations include Dynamic Layer Norm for speech recognition \cite{kimDynamicLayerNormalization2017}, Conditional Instance Norm \cite{ghiasiExploringStructureRealtime2017a}, Adaptive Instance Norm \cite{huangArbitraryStyleTransfer2017}, and Conditional Batch Norm as used in Occupancy networks \cite{meschederOccupancyNetworksLearning2019}.

\section{Experimental setup}
\label{exper}
\paragraph{Parameterizations of $\boldsymbol{\phi_i}$ \& $\mathbf{p\boldsymbol{(}\boldsymbol{\theta}\boldsymbol{|}z\boldsymbol{)}}$}
 The following parameterizations of $\phi_i$ \& $p(\theta|\mathbf{z})$ are compared, exact parameter counts are in brackets:

\begin{enumerate}
    \item Transposed Convolutional Neural Networks (TCNN \cite{donahueAdversarialAudioSynthesis2019}) [800k]
    \item Periodic Conditional Implicit Neural Representations (PCINR \cite{chanPiGANPeriodicImplicit2020}, 256 hidden units, 8 layers) [790k]
    \item Wide Periodic Conditional Implicit Neural Representations (PCINR Wide, 380 hidden units, 4 layers) [850k]
    \item Periodic Conditional Implicit Neural Representations with Weight Regularization (PCINR WR, 256 hidden units, 8 layers) [790k]
\end{enumerate}

For all parameterizations of $\phi_i$ \& $p(\theta|\mathbf{z})$, the network architecture is changed such that parameter counts are in the order of $10^6$. TCNN's parameter count is reduced by reducing the number of channels throughout the WaveGAN decoder network by a factor of eight. Models are trained with latent codes of size 256

\subsection{Objective function}

In our experiments we use MSE with addition of MSE between the derivative of target signals approximated with forward finite difference, and the derivative of reconstructions evaluated in the PCINRs (For TCNNs this is approximated by forward finite difference derivative of the reconstruction):

\begin{align*}
  \begin{split}
      \mathcal{L}=\frac{1}{N}\frac{1}{M}\sum_{i=1}^{N}\sum_{j=1}^{M} &\left\|\Phi(t_{j}, \mathbf{z_{i}})-\mathrm{y}_{i}(t_{j})\right\|^{2} \\
      + &\left\|\frac{\delta}{\delta t_{j}}[\Phi(t_{j}, \mathbf{z_{i}})]-\Delta{t_{j}}[\mathrm{y}_{i}(t_{j})]\right\|^{2}.
  \end{split}
\end{align*}
    
Where N is the batch size, M is the amount of sampled time coordinates per waveform and $\Delta$ is the forward finite difference. The addition of a MSE of derivatives of signals is justified by consistent training improvements observed in preliminary experiments. 


All reported results in table \ref{tab:results} are after training until convergence ($\sim$ 5k epochs).

\subsection{Weight regularization}

We use weight regularization as in the following function:

\begin{align*}
    \begin{split}
        \mathcal{L}&=\frac{1}{N}\frac{1}{M}\sum_{i=1}^{N}\sum_{j=1}^{M} \left\|\Phi(t_{j}, \mathbf{z_{i}}))-\mathrm{y}_{i}(t_{j})\right\|^{2} \\
        &+ \left\|\frac{\delta}{\delta t_{j}}[\Phi(t_{j}, \mathbf{z_{i}})]-\Delta_{t_{j}}[\mathrm{y}_{i}(t_{j})]\right\|^{2} +\frac{\lambda}{2}\|\mathbf{W}_{\phi_i}\|^{2}.
    \end{split}
\end{align*}

\noindent Note that weight regularization is \textit{only} applied to the weights of $\phi_i$, not to those of $\psi$.

\subsection{Datasets}
We consider the following datasets:

\begin{enumerate}
    \item 1 second, 1024 item, keyboard waveforms in MIDI notes [60, 64] subset of NSYNTH \cite{engelNeuralAudioSynthesis2017}. Balanced for note counts. Recorded at a sampling rate of 16kHz. 
    \item 1 second, 1024 item, keyboard, mallet and guitar waveforms in MIDI notes [24, 84] subset of NSYNTH. Balanced for note- and instrument counts. Recorded at a sampling rate of 16kHz. 
\end{enumerate}

Preliminary tests of baseline parameterizations of $\phi_i$ and $p(\theta|\mathbf{z})$ in the specific generative framework described in section \ref{sec:fromal_problem} showed that $\phi_i$ and $p(\theta|\mathbf{z})$ with a reasonable amount of parameters were uncapable of reconstructing large datasets\footnote{This depends on many factors including the generative framework, how its parameterized, latent embedding size and dataset uniformity}. Thus, we set the dataset size for the main experiments to be challenging, but not infeasible based on preliminary experiments.

We selected a dataset subset with the smallest range of subsequent notes within one instrument family cumulating to 1024 items. This turned out to be the keyboard instrument family, MIDI note 60 (C4, 261.63Hz) to 64 (E4, 329.63Hz). The dataset was sampled balancing for note counts. We refer to this dataset as NSYNTH Keyboard.

We selected the other split of NSYNTH to contrast with NSYNTH Keyboard in pitch- and timbral diversity. Furthermore, we selected MIDI notes 24 (C1, 32.70Hz) to 84 (C6, 1046.50Hz). We selected three instruments families to introduce more timbral diversity: keyboard, mallet and guitar. The dataset was sampled balancing for note-  and instrument family counts. We refer to this dataset as NSYNTH Diverse.

\subsection{Optimizer}
Modern neural networks are typically trained with first-order gradient methods, which can be broadly categorized into two branches: the accelerated stochastic gradient descent family such as SGD with momentum and the adaptive learning rate methods, such as Adam. SGD methods use a global learning rate for all parameters, while adaptive methods compute an individual learning rate for each parameter. Compared to the SGD family, adaptive methods typically converge fast in the early training phases, but have poor generalization performance. In our baseline decoder comparison we compare Adam \cite{kingmaAdamMethodStochastic2017} and Adabelief \cite{zhuangAdaBeliefOptimizerAdapting2020}. AdaBelief consists of the same algorithm as Adam, but also considers curvature information by scaling update directions by the change in gradient. Preliminary experiments showed consistent, favourable results for Adabelief in all setups.

\subsection{Hyperparameter search}
In all experiments we optimize PCINR hyperparameters activation scaling first and activation scaling hidden. For every combination of dataset $D$, architecture and latent embedding inference method. We optimize these hyperparameters by running respective experiments in a short training regime of 200 epochs, using Bayesian Search \cite{snoekPracticalBayesianOptimization2012} with Hyperband early stopping \cite{liHyperbandNovelBanditBased2018} to guide the search. Preliminary experiments indicated that optimal activation scaling or coordinate multiplier values are very sensitive to many architectural- and data characteristics. Thus, these values are swept using the described procedure in all experiments. 

\section{Evaluation}

\label{sec:evaluation}
Capturing the perceptual fidelity of audio reconstructions in a metric is not straightforward. Many audio-to-audio distances exist, each with different sensitivities, indicating the complexity of the problem. 

For this work, we decided to start with an extensive selection of metrics based on recent and established research in audio reconstruction and generation and implementation availability. Then, we select metrics for representing background noise presence and overall quality based on correlation with a small sample of human judgements of reconstructions.

\paragraph{Considered metrics}

We compared the following metrics:

\begin{enumerate}
    \item time-domain MSE
    \item time-domain derivative MSE
    \item time-domain MSE + derivative MSE
    \item Signal-to-Noise Ratio (SNR)
    \item Segmented SNR (SegSNR)
    \item discrete Fourier transform (DFT) magnitude MSE
    \item DFT magnitude wasserstein distance
    \item DFT angular phase MSE
    \item DFT magnitude pos/neg difference
    \item Log-spectral distance (LSD) \cite{grayDistanceMeasuresSpeech1976}
    \item Multi resolution short-time Fourier transform (STFT) MSE
    \item Multi resolution STFT wasserstein distance
    \item Multi resolution STFT pos/neg difference
    \item CSIG, CBAK, COVL \cite{huEvaluationObjectiveQuality2008}
    \item PESQ \cite{rixPerceptualEvaluationSpeech2001}
    \item FAD \cite{kilgourFrEchetAudio2019b}
    \item CDPAM \cite{manochaCDPAMContrastiveLearning2021}
\end{enumerate}

\paragraph{Chosen metrics}
We use Multi resolution STFT MSE for representing background noise presence and CDPAM for overall quality.

Multi resolution STFT MSE is calculated by averaging STFT magnitude MSE's as shown in the equation with hamming windowing for N=4 window sizes: \{400, 800, 1600 3200\}:

\begin{align*}
    \frac{1}{N} \sum_{i}^{N} \||S T F T_{i}(X_i)|- |S T F T_{i}(\{\phi_i(t_{j)}\}_{j=1}^{M})|\|^{2}.
\end{align*}

CDPAM is a contrastive learning based perceptual audio similarity metric parameterized as a deep neural network. It is trained on a dataset of audio similarity judgements based on triplet comparisons, asking subjects: “Is A or B closer to reference C?”.


\section{Discussion and Future Work}
\label{future}

In all evaluated setups latent representations $\mathbf{z}$ are optimized per audio sample. We do not enforce any distribution on the latent space, nor any other method to improve generative capabilities. Thus, the current implementations allow timbre interpolation, but these interpolations contain artifacts. It should also be noted that the current implementations are not built to do conditional generation (e.g. note- or text conditioned). Unseen samples can be encoded and synthesized by iteratively optimizing latent representations $\mathbf{z}$.
 
PCINR gains are largest for MSE, which is optimized directly by the loss function during training. However, this does not translate to strong gains in perceptual metrics, showing that PCINRs do not contain a strong oscillatory bias and suggesting they could benefit significantly from spectral or perceptual objective functions.

We experimented with several spectral and perceptual objective functions, but none resulted in improvements. We argue this is caused by inconsistent gradients for near identical inputs and note three mutually reinforcing sources for this. Firstly, because the inputs of PCINRs are one-dimensional coordinates, they process relatively many close to identical inputs. Secondly, the loss landscapes of perceptual and spectral objective functions are locally inconsistent \cite{turianSorryYourLoss2020}. Finally, periodic nonlinearities with high scaling factors contain a high density of stationary points, which cause locally inconsistent signal propagation (which also stimulates noise in output signals). 

Taking the inferior label noise robustness observed in PCINRs into account, methods for dealing with noisy loss signals could be another promising avenue. This would allow the usage of spectral and perceptual reconstruction losses, which correlate much better with perceptual qualities than MSE. Combined with the expressivity of sinusoidal representation networks conditioned using feature wise linear modulation this could greatly improve perceptual qualities of represented distributions. One might hope that gradient clipping can also aid in mitigating the detrimental effects of locally inconsistent loss functions on learning. However, it has been proven that in classification problems standard gradient clipping does not \textit{in general} provide robustness to label noise \cite{menonCANGRADIENTCLIPPING2020}.

The compositional nature of PCINRs makes it difficult to analyze representation characteristics in a signal processing context.  Multiplicative filter networks \cite{fathonyMULTIPLICATIVEFILTERNETWORKS2021} (MFNs) can be viewed as linear function approximators over an exponential number of Fourier or Gabor basis functions. This establishes a connection of the network architecture with the traditional Fourier and Gabor wavelet transforms, which are extensively studied in literature and widely used in many application domains, especially audio.  

Comparisons with MFNs would be useful to indicate if the compositional depth of PCINRs form a substantial benefit. In the experiments reported by the authors, MFNs show similar performance as SIRENs in parameter equal experiments, and larger gains in performance when increasing network depth and width. We indeed found PCINRs to lack in terms of scaling gains. Combined with the intuition that Fourier-based MFNs contain a strong oscillatory bias, the architecture is a promising avenue for future research in the area of implicit neural audio synthesis. 

Another idea which looks promising in the light of our results, still unexplored in current INR literature, is that of splitting the classically fully-connected MLP parametrizing INR’s in multiple parallel subnetworks. This could reduce the propagation of noise induced at stationary points and cancel noise in output by aggre- gating multiple independent noise sources. This argumentation is supported by multiple research directions.

Assuming similar principles are shared between channels in convolutional neural networks and subnetworks in INRs, several publications \cite{liVisualizingLossLandscape2018, zhangDeepNeuralNetworks2018} show that increasing the amount of kernels in Resnets has profound effects on reducing chaotic behaviour of loss landscapes.

The concept of combining the outputs of parallel subnetworks also has clear analogies to the field of ensemble methods, which could provide further theoretical foundations relating to this idea. For example, recent work of Fort et al.\cite{fortDeepEnsemblesLoss2020} shows how different randomly initialized CNN’s trained for classification problems tend to explore diverse modes in function space, resulting in improved robustness. We found that averaging waveform reproductions of independently trained conditional INR’s worked well in mitigating noisy components, indicating that similar principles might apply.

\section{Broader Impact}

The development of neural audio synthesis technologies has the potential to negatively impact musicians who make a living as professional performers. To what extent this is already occurring is
unclear, however, it is consistent with a trend of increased access to music production methods
that had previously been reserved primarily for professional recording studios.

\printbibliography 




\end{document}